\documentclass[reprint,
groupedaddress,
showpacs,preprintnumbers,
amsmath,amssymb,
aps,
prb
]{revtex4-1}
\usepackage{graphicx}
\usepackage{dcolumn}
\usepackage{bm}
\usepackage{mathrsfs}
\usepackage{color}
\begin{document}
\title{New approach to theory of tunneling spectroscopy in unconventional superconductors }


\author{A.\,V.~Burmistrova}
\affiliation{Lomonosov Moscow State University Skobeltsyn Institute of Nuclear Physics, 1(2), Leninskie gory, GSP-1, Moscow 119991, Russian Federation}

\author{ I.\,A.~Devyatov}
\email[]{igor-devyatov@yandex.ru}
\affiliation{Lomonosov Moscow State University Skobeltsyn Institute of Nuclear Physics, 1(2),  Leninskie gory, GSP-1, Moscow 119991, Russian Federation}

\author{Alexander A. Golubov}
\affiliation{Faculty of Science and Technology and MESA+ Institute of Nanotechnology,
University of Twente, 7500 AE, Enschede, The Netherlands}

\author{Keiji Yada}
\affiliation{Department of Applied Physics, Nagoya University, Nagoya 464-8603, Japan}

\author{Yukio Tanaka}
\affiliation{Department of Applied Physics, Nagoya University, Nagoya 464-8603, Japan}


\date{\today}

\begin{abstract}
We have derived new boundary conditions on wave function at the
normal metal / superconductor (NS) interface beyond effective mass approximation. These conditions are based on tight-binding approach
and enable one to formulate quantitative model for tunneling spectroscopy of superconductors with  complex
non-parabolic energy spectra.  The model is applied to superconductors with unconventional pairing and with
multiband electronic structure. In the case of single band unconventional superconductors this model provides known
conductance formula (Phys. Rev. Lett. 74 3451 1995), but with generalized definition of the normal-state conductance.
Based on new boundary conditions, we have calculated conductance in normal metal / superconducting pnictide junctions
for different orientations of the NS interface with respect to the crystallographic axes of the pnictides.
The present approach provides the basis for quantitative tunneling spectroscopy in multi-orbital superconductors.

\end{abstract}

\pacs{74.20.Rp,74.70.Xa,74.45.+c,74.50.+r,74.55.+v}

\maketitle

\section{\label{sec1}Introduction}

Tunneling spectroscopy was extensively applied up to now to reveal
important features of  electronic properties of
superconductors.\cite{Wolf} It was predicted long time ago \cite{Bardeen,Cohen} that in a tunnel junction
between a normal metal and a metallic BCS superconductor (NS)
the differential conductance $dI/dV$ is expressed by the bulk density of states.
Later, Blonder, Tinkham and Klapwijk (BTK) formulated the model for $dI/dV$
in NS junctions with arbitrary barrier transparency by solving Bogoliubov-de Gennes (BdG) equation
and explicit calculation of the Andreev and normal reflections coefficients.\cite{btk}
It was shown that in the regime of low transparency of NS interface the resulting
$dI/dV$ corresponds to the energy spectrum of local density of states
in a bulk superconductor.
On the other hand, $dI/dV$ in high transparency limit is controlled by the Andreev reflection.
The original formulation of the BTK theory has provided the basis for
tunneling spectroscopy and has been widely applied for many conventional
superconducting junctions where the symmetry of superconductor
is conventional spin-singlet $s$-wave.
The magnitude of energy gap for conventional $s$-wave
superconductor has been determined with high accuracy.
However, the effects of anisotropy of pair potential in
$d$-wave pairing were not included in the original BTK formula. Therefore,
after the discovery of high-$T_{C}$ cuprates, the extension of BTK
formula become really needed. Such an extension to unconventional superconductors
has been formulated in Refs. \onlinecite{Bruder,tanaka1}.
It was shown that tunneling conductance of unconventional superconducting junctions
is not always expressed by the bulk density of states due to the  presence of
surface Andreev bound states (ABS).\cite{tanaka1,kashiwaya00}
It has been revealed by a theory of tunneling spectroscopy
of spin-singlet $d$-wave superconductor \cite{tanaka1}
that zero bias conductance peak (ZBCP) stems from the surface zero energy ABS
frequently observed in the experiments of high-$T_{C}$ cuprates.\cite{Hu}
This ABS has a flat band dispersion and originates from the
sign change of the pair potential on the Fermi surface.
Due to this flat dispersion, ZBCP ubiquitously emerges in actual experiments.
\cite{tanaka1,kashiwaya00,Lofwander}
A number of anomalous quantum phenomena like non-monotonic temperature dependence of
Josephson current in high $T_{C}$ cuprate stem from this ABS.
\cite{tanaka96,Barash,tanaka97}
Further, theory of tunneling spectroscopy of  normal metal /spin-triplet $p$-wave
superconductor junctions has been formulated \cite{YTK97,YTK98,Honerkamp}
stimulating by the discovery of Sr$_{2}$RuO$_{4}$.\cite{mae}
It was shown that the line shape of a tunneling conductance in chiral $p$-wave superconducting
junctions has a broad ZBCP due to the ABS with linear dispersion.\cite{MS99,FMS01,Kashiwaya11}

At present, to clarify the pairing mechanism of
iron-based pnictides is one of hot topics in condensed matter physics.
Just after the discovery of superconductivity in pnictides,
$s_{\pm}$ symmetry of the pair potential (order parameter)
has been proposed,\cite{maz1,Kuroki} where
pair potential changes sign between electron and hole Fermi pockets.
The glue of this pairing is provided by spin-fluctuations, which typically appear in strongly correlated
superconductors.
On the other hand, $s_{++}$-wave pairing symmetry due to orbital fluctuations
has been proposed as another candidate of superconducting pairing.\cite{Kontani}
The possibility of the inter-orbital pairing was also discussed.\cite{mor}
Since iron pnictides are multi-orbital systems with multiple orbital systems,
a theory of tunneling spectroscopy applicable to multi-orbital superconductors
is strongly needed.
Besides pnictides, there are many new unconventional superconductors with multiple Fermi surfaces
such as doped topological insulator Cu$_x$Bi$_2$Se$_3$.
\cite{hor10,sasaki11,koren11,kirzhner12,yang12,fu10,hao11,hsieh12,yamakage12}


However, formulation of microscopic theory of tunneling spectroscopy in
multi-band superconductors is highly nontrivial task.
The most crucial point is the boundary conditions on
wave functions at the NS interface.
Ara\'{u}jo and Sacramento  have presented a new way to describe
boundary conditions between single-band normal metal and
multi-band superconducting systems
using phenomenological approach based on analogy between quantum waveguide theory and interband scattering.\cite{sacr}
This idea was applied to actual pnictide junctions,\cite{sacr,dev1}
but the basis of this theory is not fully microscopic since
interband and intervalley scattering effect is not fully taken into account.
Other theories devoted to the study of coherent transport in junctions of iron
pnictides are also phenomenological.\cite{lind,gol,rom,kar1,kar2,chen,ber}
The formulation of boundary conditions on the wave function in multi-band systems has not
become clear up to now.

In this paper, based on equations of tight-binding model, we obtain the boundary
conditions on the wave functions
for the contact between a normal metal and a multiband superconductor.
Up to now, tight-binding approach has been used for the
study of ABS in various superconductors.
\cite{Tanuma98,TanumaJ99,Tanuma99,
Tanuma2001,Tanuma2002,Tanuma2003,Onari2009}
Here, we obtain boundary conditions beyond the effective mass approximation
in order to take into account  complex nonparabolic and anisotropic spectrum of energy band in the normal state and unconventional pairing in multi-band
superconducting systems.
The obtained  boundary  conditions provide an extension of  tight-binding approach by Zhu and  Kroemer \cite{krom} to a superconducting case.
The approach \cite{krom} is physically transparent, since the only assumption  is the prolongation of solutions of tight-banding model
on one additional site to the left (right) sides of an interface.
We apply the derived boundary conditions to the calculation of charge conductance between
a normal metal and an iron pnictide superconductor for different misorientation angles between crystallographic axes of a pnictide superconductor and the interface.
Application of our theory to single band unconventional superconducting junctions allows one to
reproduce preexisting formula of tunneling spectroscopy of
unconventional superconductors,  where the transparency of the junction
in the  normal state has a different momentum dependence. Brief account of some results of this paper is given in Ref. \onlinecite{bcrus}.

\section{\label{sec2}One-dimensional model}

In order to understand the essence of new boundary conditions,
we first consider one-dimensional model of
normal metal and spin-singlet $s$-wave superconductor junction.
We use a model Hamiltonian $H$ of the $1D$ chain of atoms, whose
electronic states are described in the tight-binding approximation,
where Cooper pair is formed on the same site:

\begin{equation}
H=H_N+H_S+H_I,
\label{ham}
\end{equation}
\begin{equation}
H_N = \sum_{n\le0,\sigma}\left[t'\left( a_{\sigma,n-1}^\dag a_{\sigma,n}+h.c. \right) - \mu_N a_{\sigma,n}^\dag a_{\sigma,n} \right],
\end{equation}
\begin{eqnarray}
H_S &=& \sum_{n\ge1,\sigma}\left[t\left( a_{\sigma,n}^\dag a_{\sigma,n+1}+h.c. \right) - \mu_S a_{\sigma,n}^\dag a_{\sigma,n} \right]\nonumber\\
&&-{\sum_{n}\left[  \Delta a_{\uparrow,n}^\dag a_{\downarrow,n}^\dag +h.c.  \right]  },
\end{eqnarray}
\begin{gather}
H_I= \gamma\left( a_{\sigma,0}^\dag a_{\sigma,1}+h.c. \right),
\end{gather}

\noindent with creation (annihilation) operator
$a_{\sigma,n}^\dag \left(a_{\sigma,n}\right)$   of an electron with spin $\sigma$ on $n$ site,
and pair potential $\Delta$.
$t'$ ($t$) and $\mu_S$ ($\mu_N$) are hopping and chemical potential in normal metal (superconductor), respectively.
$H_N$, $H_S$ and $H_I$ are hamiltonian in the normal metal (N), in the superconductor (S) and at the interface, respectively.
Eq. (\ref{ham}) is diagonalized
by introducing the following canonical transformation:

\begin{equation}
a_{\sigma,n}=\sum_{\nu}\left[ u_{\nu,n} \alpha_{\nu,\sigma} +{\rm sgn}(\sigma)v_{\nu,n}^\ast \alpha_{\nu,-\sigma}^\dag   \right] ,\label{bog}
\end{equation}

\noindent which is a generalization of the Bogoliubov transformation \cite{tink} to the case of a discrete lattice.
In Eq. (\ref{bog}) $\alpha_{\nu,-\sigma}^\dag(\alpha_{\nu,\sigma})$ are
operators of creation (annihilation) of quasiparticles satisfying Fermi
anticommutation relations, and $u_{\nu,n}$, $v_{\nu,n}$ are wave functions in BdG equation.
The discrete version of BdG equations for wave functions  $u_{\nu,n}$, $v_{\nu,n}$ has the form:

\begin{equation}
\left\{
\begin{aligned}
&t_n u_{\nu,n+1}+t_{n-1} u_{\nu,n-1} - \mu_n u_{\nu,n}\\
&+\Delta_n v_{\nu,n}=\varepsilon_{\nu}u_{\nu,n},\\
&t_n v_{\nu,n+1}+t_{n-1}v_{\nu,n-1} - \mu_n u_{\nu,n}\\
&-\Delta^\ast_n u_{\nu,n} =-\varepsilon_{\nu}v_{\nu,n},
\end{aligned}
\right. \label{bogeq}
\end{equation}
where $\mu_n=\mu_N$ ($\mu_S$) for $n\le0$ ($n\ge1$),
and $t_n=t'$, $\gamma$ and $t$ for $n\le-1$, $n=0$ and $n\ge 1$,
respectively.
Basically, Eq. (\ref{bogeq}), together with the corresponding self-consistent equations for the pair potential $\Delta_n$, provides
the description for any spatially-inhomogeneous problem with an arbitrary set of hopping parameters between sites $t_n$.
However, this problem can be solved only numerically.
In order to formulate a relevant simplified model for an NS junction which allows analytical solution,
we assume that there is no inhomogeneity of pair potential in a superconductor.
That means, $\Delta_n=\Delta$ for $n\ge1$ and $\Delta_n=0$ for $n\le0$. The structure under consideration is depicted in Fig. 1.

\begin{figure}[t!]
\centerline{\includegraphics[width=8cm]{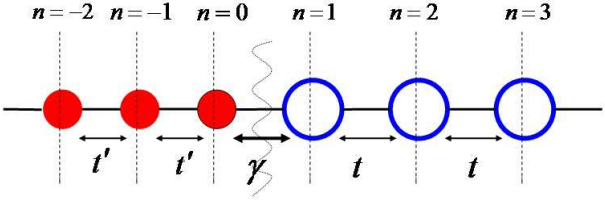}}
\caption{Schematic illustration of one-dimensional model.
Left region (red filled circles) corresponds to that of normal metal with hopping parameter $t^{'}$, right region (blue circles) corresponds to that of superconductor
with hopping parameter $t$. The hopping parameter
at  NS boundary is given  by $\gamma$.}
\label{Fig1}
\end{figure}

First,  we consider the conductance in a normal metal / normal metal junction by setting $\Delta=0$ in a superconducting region.
The electron with energy $E(=\varepsilon_{\nu})$ is injected from the left side
and it is scattered at the interface.
Then the wave functions $u_{\nu,n}$ for the left side $\Phi_n(=u_{\nu,n})$ with $n\le0$ and the right side $\Psi_n(=u_{\nu,n})$ with $n\ge1$ are given by
\begin{equation}
\left\{
\begin{aligned}
&\Phi_n=\exp(iqnl)+b~\exp(-iqnl),\\
&\Psi_n=c~\exp(iknl),
\end{aligned}
\right.\label{var1-0}
\end{equation}
where $l$ is the lattice constant in a normal metal and a superconductor (for clarity, we consider them to be equal,
but this restriction is not required.\cite{krom})
The first and the second term of $\Phi_n$ represent an incident and a normal reflected waves, respectively.
$\Psi_n$ corresponds to a transmitted wave.
Here, $q$ and $k$ are determined by the equation
$2t'\cos(ql)=\mu_N + E$
and $2t\cos(kl)=\mu_S +E$
with $q, k\ge0$, respectively.
The coefficients $b$ and $c$ are determined by the boundary conditions.
These boundary conditions were proposed by Zhu and Kroemer.\cite{krom}
Their method is not limited by the assumption of
parabolic single-particle excitation spectrum based on the electronic effective-mass concept.
In their idea, one can obtain the boundary conditions by the shift of the location of the boundary.
If we shift the boundary to the right, we obtain the Shr\"{o}dinger equation
\begin{equation}
E\Phi_0=-\mu_N\Phi_0+t'\Phi_{-1}+t'\Phi_{1},\label{eq:1}
\end{equation}
where $\Phi_{1}$ is obtained by the natural extension of $\Phi_n$ ($n\le0$) given in Eq. (\ref{var1-0}).
The Shr\"{o}dinger equation without the shift of the boundary is given by
\begin{equation}
E\Phi_0=-\mu_N\Phi_0+t'\Phi_{-1}+\gamma\Psi_{1}.\label{eq:2}
\end{equation}
By subtracting Eq. (\ref{eq:1}) from Eq. (\ref{eq:2}), we obtain the boundary condition
\begin{equation}
t'\Phi_1=\gamma \Psi_1\label{bc1-1}.
\end{equation}
Similarly, if we shift the boundary to the left, we obtain the boundary condition
\begin{equation}
\gamma \Phi_0=t\Psi_0.\label{bc1-2}
\end{equation}
Using the boundary conditions Eqs. (\ref{bc1-1}) and (\ref{bc1-2}) and the wave functions Eq. (\ref {var1-0}),  one can obtain $b,c$ given in  Eq. (\ref {var1-0}):

\[
b=\frac{\sigma_{1}\exp(iql)-\exp(ikl)}
{\exp(ikl) -\sigma_{1} \exp(-iql)},
\]

\begin{equation}
c=\gamma (1+b)/t,
\end{equation}

\noindent with $\sigma_{1}=t t'/ \gamma^{2}$,
and expression for the transparency $\sigma_{N}$ at the interface:

\begin{equation}
\sigma_{N}(k,q)
=1 - \mid b \mid^{2}
=\frac{ 2\sigma_{1}[\cos[(q-k)l]
-\cos[(q+k)l]]}
{1 + \sigma_{1}^{2} - 2\sigma_{1}\cos[(q+k)l] }.
\label{condn}
\end{equation}

The boundary conditions Eqs. (\ref{bc1-1}) and (\ref{bc1-2}) provide the conservation of probability flow $J$ across the interface:

\begin{equation}
J_{n\le-1}=\frac{2t'}{\hbar}{\rm Im}(\Phi^\ast_{n+1}\Phi_n)=J_{n>1}=\frac{2t}{\hbar}{\rm Im}(\Psi^\ast_{n+1}\Psi_n).
\label{f1}
\end{equation}

After introducing finite difference derivatives in the following form:
$\psi'_1=(\Psi_1-\Psi_0)/l ,
\psi'_2=(\Phi_1-\Phi_0)/l
$, the boundary conditions Eqs. (\ref{bc1-1}) and (\ref{bc1-2}) lead to usual boundary conditions,\cite{laikh} obtained early in the continuum limit.
It is necessary to note that these boundary conditions, written in the form of finite differences, coincide with the mostly used Harrison's boundary conditions \cite{harrison} only for the case $\sigma_{1}=1$. This feature of discrete boundary conditions Eqs. (\ref{bc1-1}) and (\ref{bc1-2}) were also mentioned in Ref. \onlinecite{krom}.

By extending the method  \cite{krom} to the case of superconducting junctions,
one can obtain from Eq. (\ref{bogeq}) the following four boundary conditions at the interface between
normal metal and spin-singlet $s$-wave superconductor junctions (see Fig. 1):

\begin{equation}
\left\{
\begin{aligned}
&t'\Phi_1=\gamma \Psi_1,\\
&t'\bar{\Phi}_1=\gamma \bar{\Psi}_1,\\
&\gamma \Phi_0=t\Psi_0,\\
&\gamma \bar{\Phi}_0=t\bar{\Psi}_0,
\end{aligned}
\right.\label{bc2}
\end{equation}
where $\Psi_n (\Phi_n)$ and $\bar{\Psi}_n (\bar{\Phi}_n)$ are the wave functions $u_{\nu,n}$ and $v_{\nu,n}$ for electron and hole in S (N), respectively.
These wave functions are
given by
\begin{equation}
\left\{
\begin{aligned}
&\Phi_n=\exp(iqnl)+b~\exp(-iqnl),\\
&\bar{\Phi}_n=a~\exp(i\widetilde{q}nl),\\
&\Psi_n=c~u\exp(iknl)+d~v~\exp(-i\widetilde{k}nl),\\
&\bar{\Psi}_n=c~v\exp(iknl)+d~u~\exp(-i\widetilde{k}nl).
\end{aligned}
\right.\label{var1}
\end{equation}

The wave functions of a normal metal and a superconductor contain four unknowns $a$, $b$, $c$, $d$
 describing the Andreev and normal reflected waves in a normal metal,
and two transmitted waves in the superconductor ($c$ and $d$),  where $c$ ($d$) corresponds to transmission process by electron-like (hole-like) quasiparticles.
These four unknowns ($a$, $b$, $c$, $d$) are uniquely defined by four boundary conditions Eq. (\ref{bc2}). In Eq. (\ref{var1}),   $q, \widetilde{q}$ ($k, \widetilde{k}$) are wave vectors in normal metal (superconductor), corresponding
to the energy  $E$.

Although $q$ and $ \widetilde{q}$ are real numbers,
$k$ and $\tilde{k}$ become complex when
$\mid E \mid < |\Delta|$  is satisfied.
One can show that obtained  wave functions  provide the conservation of the probability flow by postulating boundary
conditions Eq. (\ref{bc2}).
The expression for the probability flow on a discrete lattice (Fig. 1) follows from the BdG equations  on the sites of the crystal lattice Eq. (\ref{bogeq}):

\begin{equation}
J_{s}=\frac{2}{\hbar}{\rm Im}(t\Psi^\ast_{n+1}\Psi_n-t\bar{\Psi}^\ast_{n+1}\bar{\Psi}_n)  .\label{f2}
\end{equation}

It is necessary to note that the condition for the conservation of probability flow at the interface between N and S, having the form of discrete sums (differences) in the crystal lattice (Eqs. (\ref{f1}), (\ref{f2})), can be written in a quadratic form in terms of the probability amplitudes to be in states with wave vectors $q,-q,\widetilde{q},k,-\widetilde{k}$ multiplied on the group velocities in these states:

\begin{gather}
\frac{\partial \varepsilon_n}{\partial p}|_{p=q}-|a|^2\frac{\partial \varepsilon_n}{\partial p}|_{p=\widetilde{q}}+ |b|^2\frac{\partial \varepsilon_n}{\partial p}|_{p=-q}=|c|^2\frac{\partial \varepsilon_s}{\partial p}|_{p=k}
\notag\\+ |d|^2\frac{\partial \varepsilon_s}{\partial p}|_{p=-\widetilde{k}} \label{f3}.
\end{gather}

\noindent Eq. (\ref{f3}) is similar to the corresponding expression in Ref. \onlinecite{btk}.

The above consideration of the tight-binding approximation of one-dimensional model of the NS junctions corresponds to equilibrium situation with
zero voltage at the boundary $V = 0$. However, it can be generalized to the case of a finite voltage $V \neq 0 $ on the microconstriction of atomic sizes
with a characteristic size much smaller than the elastic  $l_{el}$ and inelastic $l_{in}$ characteristic mean free paths. In such pure microconstriction electron transport can be considered as a transport on the independent transverse modes.
The current flowing through one mode is determined by the difference between the incoming $ f ^ {\to} (E) $ and outgoing $ f ^ {\gets} (E) $ flows of electrons in the normal metal:\cite{btk}

\begin{equation}
I(V)=\eta_1\int\{f^{\to}(E)-f^{\gets}(E)\}dE,
\label{current}
\end{equation}

\noindent where $f^{\to}(E)=f_0(E-eV)$,   $f_0(E)$ - the equilibrium Fermi distribution, $\eta_1=e/(\pi \hbar)$, and

\begin{gather}
f^{\gets}(E)=A(E)(1-f^{\to}(-E))+B(E)f^{\to}(E)\notag\\
+(C(E)+D(E))f_0(E).
\label{fout}
\end{gather}

\noindent In Eq. (\ref{fout}) $A(E)$, $B(E)$,
$C(E)$ and $D(E)$ are probabilities of the Andreev reflection, normal reflection, transmission as a electron-like quasiparticle and
as a hole-like quasiparticle, respectively.
The probabilities $A,B,C,D$ in Eq. (\ref{fout}) are calculated from the boundary conditions Eq. (\ref{bc2}) and the expressions for the probability flow Eqs.  (\ref{f1}), (\ref{f2}).
In calculating the probabilities $ A, B, C, D $ the incoming quasiparticle states must be normalized
so that the probability flow in these states, described by the Eqs. (\ref {f1}), (\ref {f2}), is equal to unity.
This normalization provides a thermodynamic equilibrium in the absence of voltage $ V = 0 $ on the NS junction.

For the majority of superconductors,  the magnitudes of $\Delta$ and $E$ are much smaller than those of $t$ and $t'$ and the following conditions
\begin{equation}
|\Delta/t|\ll 1,|\Delta/t'|\ll 1
\label{qc}
\end{equation}
are satisfied. This is the so-called quasiclassical approximation.
Then, the relations $q \simeq \widetilde{q} \simeq q_{0}$ and
$k \simeq \widetilde{k} \simeq k_{0}$ are satisfied, where
$k_{0}$ and $q_{0}$ are momenta at the Fermi surface satisfying
$2t'\cos(q_{0}l)=\mu_{N}$ and $2t\cos(k_{0}l)=\mu_{S}$.
The resulting amplitudes $a$ and $b$ are given by

\begin{gather}
a=\frac{2\sigma_{1}\Gamma (\cos[(q_{0}-k_{0})l]
-\cos[(q_{0}+k_{0})l])}{\Lambda},\notag\\
b=\frac{(1 - \frac{\delta \sigma_{1}}{\tilde{\delta}})
(1 - \frac{\sigma_{1}}{\delta \tilde{\delta}})}{\Lambda}
\label{ab}
\end{gather}

\noindent with

$\Gamma=\Delta/(E + \sqrt{E^{2}-\Delta^{2}})$,
$\exp(iq_{0}l)=\delta$,
$\exp(ik_{0}l)=\tilde{\delta}$

\noindent and
\[
\Lambda
= -(1 - \sigma_{1}\delta \tilde{\delta})
(1 - \sigma_{1}\frac{1}{\delta \tilde{\delta}})
[1 - (1 - \sigma_{N}(k_0,q_0)\Gamma^{2})],
\]
where
$\sigma_{N}(k_0,q_0)$ is defined by Eq. (\ref {condn})

Within this approximation,
we can reproduce  the BTK result \cite{btk}

\begin{equation}
I(V)=\eta_1\int\{f_0(E-eV)-f_0(E)\}\sigma(E)dE,
\label{current2}
\end{equation}

\noindent where

\begin{eqnarray}
\sigma(E)&=&
1 + \mid a \mid^{2} - \mid b \mid^{2}\notag\\
&=&
\frac{ \sigma_{N}[1 + \sigma_{N} \mid \Gamma \mid^{2}
+ (\sigma_{N} -1) \mid \Gamma \mid^{4}] }
{\mid 1 - (1 - \sigma_{N}) \Gamma^{2} \mid ^{2}}.
\label{conds}
\end{eqnarray}

\noindent This is the well-known formula \cite{btk} with
extended definition of the transparency
$\sigma_{N}$ (see Eq. (\ref {condn})) at  the N/S interface.

In this section, we have studied one-dimensional model
with spin-singlet $s$-wave superconductor
as the simplest case.
However, the case of the contact between a normal metal and a superconductor with anisotropic sign-changing pair potential on the Fermi surface
is of most interest. The next two sections will be devoted to the consideration of this situation.

\section{\label{sec3}two-dimensional model for the contact of a normal metal and d-wave single band superconductor}

In this section, we extend our approach to unconventional superconductor.
We show one typical example of two-dimensional
lattice model of unconventional superconductor.
BdG equations on sites of the lattice of the $d$-wave superconductor in the $x-y$ plane have the following form:

\begin{equation}
\left\{
\begin{aligned}
&t_1( \Psi_{n+1,m}+ \Psi_{n-1,m})\\&+t_2( \Psi_{n,m+1}+ \Psi_{n,m-1}) - \mu_{S}\Psi_{n,m}\\& +\Delta_0( \bar{\Psi}_{n+1,m}+\bar{\Psi}_{n-1,m}-\bar{\Psi}_{n,m+1}- \bar{\Psi}_{n,m-1})\\&=\varepsilon\Psi_{n,m}, \\
&t_1( \bar{\Psi}_{n+1,m}+ \bar{\Psi}_{n-1,m})\\&+t_2( \bar{\Psi}_{n,m+1}+ \bar{\Psi}_{n,m-1})- \mu_{S}\bar{\Psi}_{n,m}\\& -\Delta_0( \Psi_{n+1,m}+\Psi_{n-1,m}-\Psi_{n,m+1}- \Psi_{n,m-1})\\&
=-\varepsilon\bar{\Psi}_{n,m},
\end{aligned}
\right.\label{bogeqd}
\end{equation}

\noindent where $t_1,t_2$ are hopping amplitudes between orbitals on sites, $n,m$ are the numbers of site in $x$- and $y$-direction, respectively. The value of $\Delta_0 $ is the amplitude of the anisotropic pair potential corresponding to the considered $d$-wave superconducting pairing: $\Delta (k) = 2\Delta_0(\cos {k_x} - \cos {k_y})$, where $k_x$ and $k_y$ is quasimomentum
perpendicular and parallel to the interface, respectively.

The boundary conditions for the contact of a normal metal and $d$-wave superconductor, described by the Eq. (\ref {bogeqd}), in the quasiclassical limit Eq. (\ref{qc})  are the same as boundary conditions Eq. (\ref {bc2}). For the case under consideration,  $\Psi_{n}$ ($\bar{\Psi}_{n}$) in boundary conditions Eq. (\ref {bc2}) means the wave function of layer $n$ of atoms of $d$-wave superconductor in the $x-y$ plane.
Due to the translational symmetry in the $y$-direction in the electron (hole) wave functions $\Psi_{n,m}$ ($\bar{\Psi}_{n,m}$) we can omit second subscript $(m)$ corresponding to the coordinate of an atom in a direction parallel to the boundary.
It should be remarked that within quasiclassical approximation,
these boundary conditions are satisfied for any type of
unconventional superconductors.

Let's consider the situation, when the misorientation angle between interface and crystallographic axes of superconductor is equal to $\pi/4$.
In this case the current through the two-dimensional pure microconstriction between a normal metal and $d$-wave superconductor is determined by the integration over the transverse quasimomentum $k_y$ of Eq. (\ref{current2}): $I_p(V)=\eta_2\int dk_y I(V,k_y)$, where $\eta_2=\Xi/2\pi$, $\Xi$ - characteristic size of microconstriction, with the following definition of   $\sigma(E)$:

\begin{equation}
\sigma(E)=
\frac{ \sigma_{N}[1 + \sigma_{N} \mid \Gamma \mid^{2}
+ (\sigma_{N} -1) \mid \Gamma \tilde{\Gamma} \mid^{2}] }
{\mid 1 - (1 - \sigma_{N}) \Gamma \tilde{\Gamma} \mid^{2}},
\end{equation}

\noindent with $\Gamma = \Delta_{+}/(E + \sqrt{E^{2} -\Delta_{+}^{2}})$
and
$\tilde{\Gamma}=\Delta_{-}/(E + \sqrt{E^{2} -\Delta_{-}^{2}})$
with $\Delta_{\pm}=\Delta(\pm k_{x},k_{y})$.

\noindent The present result is nothing but the formula of unconventional
tunneling conductance by Tanaka and Kashiwaya
\cite{tanaka1,Kashiwaya96,kashiwaya00} with generalized definition of $\sigma_{N}$, presented by Eq. (\ref {condn}).
The surface ABS is generated when the
denominator of $\sigma(E)$ becomes zero for $\sigma_{N}=0$.
This condition is given by
\[ 1=\Gamma \tilde{\Gamma}.
\]

Here, we consider $d_{xy}$-wave pairing with $\Delta_{+}= -\Delta_{-} = 4\Delta_0 \sin {k_x} \sin {k_y}$.
After simple manipulation, the energy level of surface ABS becomes
$E=0$ for any $k_{y}$ on the Fermi surface.
The dispersionless surface ABS is generated in this case.
The topological origin of this flat band ABS has been clarified recently.\cite{Sato,tanaka12}

\section{\label{sec4}two-dimensional model for the contact of a normal metal and a two band superconducting pnictide}

\subsection{\label{subsec1}Zero misorientation angle}
\begin{figure}[h]
\centerline{\includegraphics[width=8cm]{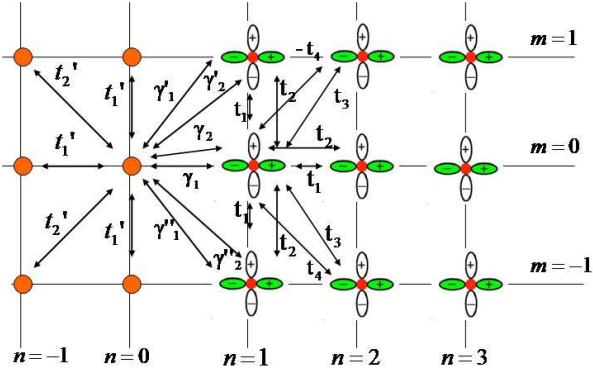}}
\caption{Boundary of normal metal / superconducting pnictide junction without misorientation.
Left region (orange circles) corresponds to the region of normal metal with hopping parameter $t'_1,t'_2$,
right region (sites with two $d$-orbitals) corresponds to the region of a superconducting pnictide with hopping parameters $t_1,t_2,t_3,t_4$.
$\gamma_1$ ($\gamma_2)$ and $\gamma'_{1}$ and $\gamma''_{1}$ ($\gamma'_{2}$ and $\gamma''_{2}$) are perpendicular and oblique hopping across the boundary to $d_{xz}(d_{yz})$-orbitals of a pnictide, respectively.
}
\label{Fig2}
\end{figure}

Consider the application of this method for the case of two-dimensional electron transport through the boundary of a normal metal and a superconducting pnictide.
In superconducting pnictides, there are two kinds of Fermi surfaces.
One is the hole-like Fermi surface around $\Gamma$-point, and the other is the electron-like Fermi surface around the zone boundary.
The minimum model to reproduce these Fermi surfaces is two-band model considering the $d_{zx}$ and $d_{yz}$ orbitals in iron 3$d$-orbitals.\cite{rag}
In this model, there are four kinds of hopping parameters $t_1$, $t_2$, $t_3$ and $t_4$.
As shown in Fig. \ref{Fig2},
$t_1$ ($t_2$) is the intra-orbital hopping between $d_{zx}$ ($d_{yz}$)-orbitals in nearest neighbor site,
$t_3$ and $t_4$ are intra-orbital and inter-orbital hopping between next nearest neighbor sites, respectively.
For the pair potential, $s_{\pm}$ model with $\Delta=4\Delta_0\cos k_x\cos k_y$
and $s_{++}$ model with $\Delta=2\Delta_0(\cos k_x+\cos k_y)+\Delta_1$ are proposed.\cite{rag,kor}
These pair potentials correspond to intra-orbital pairing and do not depend on the type of orbital.

Let us first study the case of zero misorientation angle of the crystallographic axes of a pnictide with respect to the interface as shown in Fig. 2.
The hopping perpendicular and oblique hopping between the sites on the left side and $d_{zx}$ ($d_{yz}$)-orbitals
on the  right side are described by $\gamma_1$ ($\gamma_2$) and $\gamma'_{1},\gamma''_{1},$ ($\gamma'_{2},\gamma''_{2}$), respectively.
For simplicity, we assume that
the periods of the crystal lattices in a normal metal and a pnictide are the same.
In the following calculations, we drop oblique hopping for simplicity.
The mathematical formulation of the problem and solutions of BdG equations
in the two-dimensional case are given in Appendix.

Proceeding as well as in the derivation of boundary conditions in the 1D-model Eq. (\ref{bc2}),
but taking into account independent hopping on $d_{xz}$ and $d_{yz}$-orbitals of a pnictide,
we obtain the following boundary conditions for NS junction with zero misorientation angle:

\begin{equation}
\left\{
\begin{aligned}
&t_1'\Phi_1=\gamma_1 \Psi^{\alpha}_1+\gamma_2 \Psi^{\beta}_1 ,\\
&t_1'\bar{\Phi}_1=\gamma_1 \bar{\Psi}^{\alpha}_1+\gamma_2,
\bar{\Psi}^{\beta}_1, \\
&\gamma_1
\Phi_0=(t_1+2t_3\cos k_y)\Psi^{\alpha}_{0}+2it_4\sin k_y\Psi^{\beta}_{0}\\
&+2\Delta_0\zeta(k_y) \bar{\Psi}^{\alpha}_0,\\
&\gamma_1
\bar{\Phi}_0=(t_1+2t_3\cos k_y)\bar{\Psi}^{\alpha}_{0}+2it_4\sin k_y\bar{\Psi}^{\beta}_{0}\\
&-2\Delta_0\zeta(k_y) \Psi^{\alpha}_0,\\
&\gamma_2
\Phi_0=(t_2+2t_3\cos k_y)\Psi^{\beta}_{0}+2it_4\sin k_y\Psi^{\alpha}_{0}\\
&+2\Delta_0\zeta(k_y)\bar{\Psi}^{\beta}_0,\\
&\gamma_2
\bar{\Phi}_0=(t_2+2t_3\cos k_y)\bar{\Psi}^{\beta}_{0}+2it_4\sin k_y\bar{\Psi}^{\alpha}_{0}\\
&-2\Delta_0\zeta(k_y) \Psi^{\beta}_0,
\end{aligned}
\right.\label{bc3}
\end{equation}
where $\zeta(k_y)=\cos k_y$ and 1/2 for $s_{\pm}$ and $s_{++}$ models, respectively.
Due to the translational symmetry of this structure in a direction parallel to the boundary,
$k_y$ component of the quasimomentum is conserved.
Also because of this translational symmetry in the electron (hole) wave functions $\Psi^{\alpha (\beta)}_{n,m}$ ($\bar{\Psi}^{\alpha (\beta)}_{n,m}$),
the second subscript $(m)$ corresponding to the coordinate of an atom in a direction parallel to the boundary is omitted.
The wave functions of the NS contact are defined by 6 plane waves with amplitudes $a, b, c_1, c_2, d_1, d_2$:
$a$, $b$ describe the Andreev and normal reflected waves in normal metal.
$c_1$ ($c_2$) and $d_1$ ($d_2$) describe electron-like and hole-like transmitted waves corresponding to inner (outer) Fermi surface in a superconducting pnictide, respectively:

\begin{equation}
\left\{
\begin{aligned}
&\Phi_n=\exp(iq_1nl)+b~\exp(-iq_1nl),\\
&\bar{\Phi}_n=a~\exp(iq_2nl),\\
&\Psi^{\alpha}_n=c_1u_1(k_1)\exp(ik_1nl)+c_2u_1(k_2)\exp(ik_2nl)\\&+d_1u_1(k_3)\exp(ik_3nl)+d_2u_1(k_4)\exp(ik_4nl),\\
&\Psi^{\beta}_n=c_1u_2(k_1)\exp(ik_1nl)+c_2u_2(k_2)\exp(ik_2nl)\\&+d_1u_2(k_3)\exp(ik_3nl)+d_2u_2(k_4)\exp(ik_4nl),\\
&\bar{\Psi}^{\alpha}_n=c_1v_1(k_1)\exp(ik_1nl)+c_2v_1(k_2)\exp(ik_2nl)\\&+d_1v_1(k_3)\exp(ik_3nl)+d_2v_1(k_4)\exp(ik_4nl),\\
&\bar{\Psi}^{\beta}_n=c_1v_2(k_1)\exp(ik_1nl)+c_2v_2(k_2)\exp(ik_2nl)\\&+d_1v_2(k_3)\exp(ik_3nl)+d_2v_2(k_4)\exp(ik_4nl).
\end{aligned}
\right.\label{vf3}
\end{equation}

 \begin{figure}[h]
\centerline{\includegraphics[width=9cm]{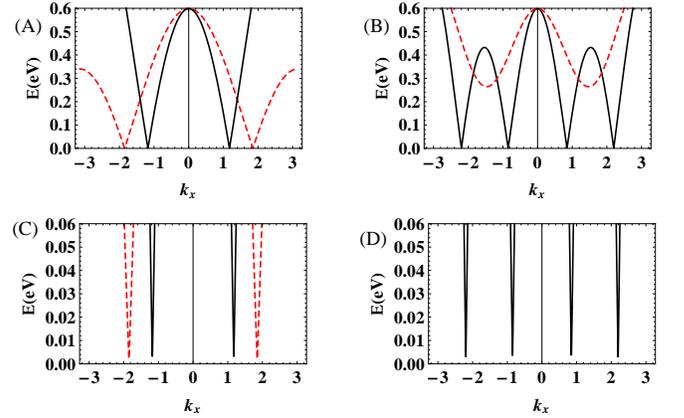}}
\caption{The excitation spectrum of a superconducting pnictide for a fixed value of $k_y$. (A) - misorientation angle is equal to 0, $k_y=0$, (B) - misorientation angle is equal to $\pi/4$, $k_y=0$, (C) corresponds to the (A), depicted on a larger scale, (D) corresponds to (B), depicted on a larger scale. Red dashed line and black solid line correspond to the two different bands.}
\label{Fig3}
\end{figure}

 \begin{figure}[h]
\centerline{\includegraphics[width=8cm]{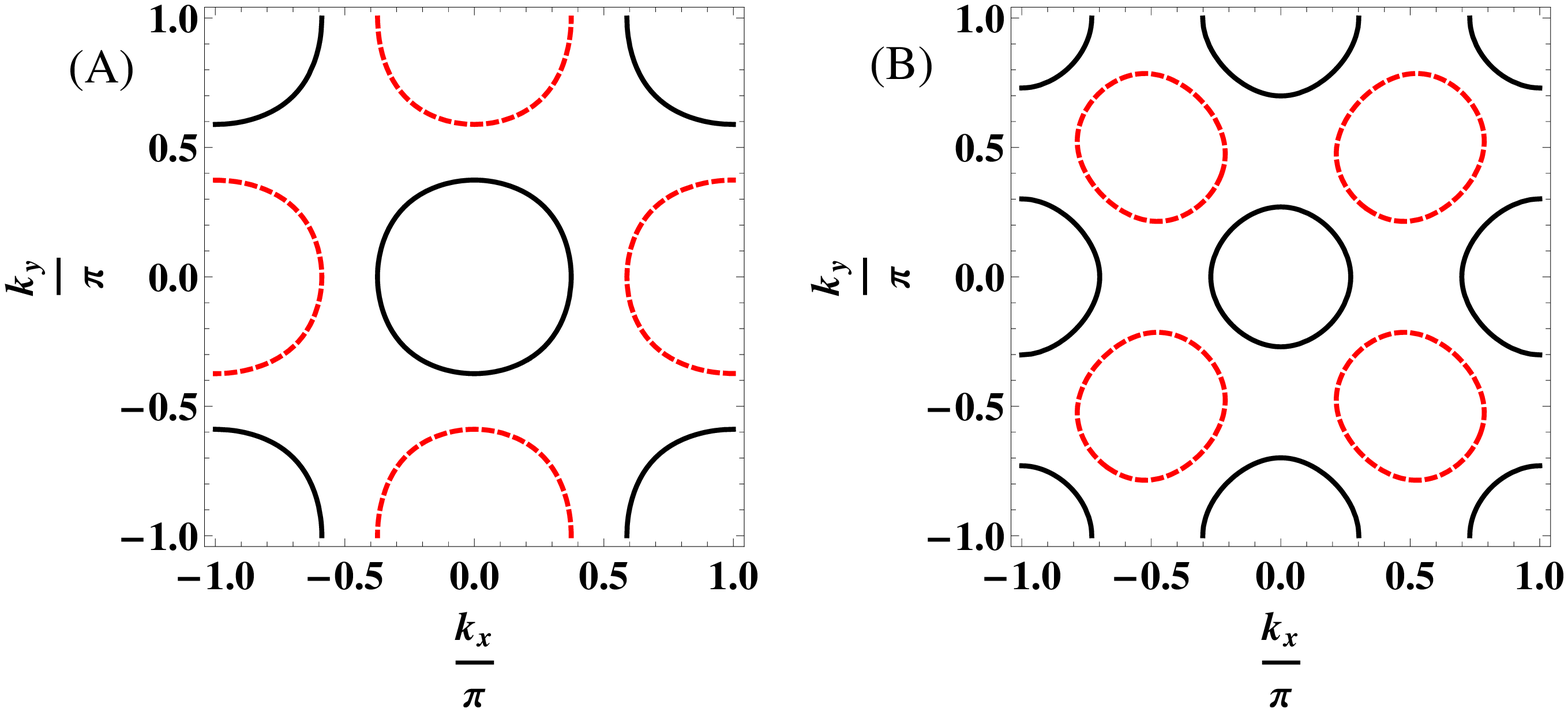}}
\caption{The Fermi surface of a pnictide in the unfolded Brillouin zone.  (A) - misorientation angle is equal to 0, (B) - misorientation angle is equal to $\pi/4$}
\label{Fig3a}
\end{figure}
\noindent
where $q_1$, $q_2$ are the wave vectors for electron and hole in normal metal with excitation energy $E$, respectively.
$k_1$ ($k_2$) and  $k_3$ ($k_4$) are the wave vectors for electron-like and hole-like quasiparticle corresponding to inner (outer) Fermi surface in pnictides.
Six coefficients $a, b, c_1, c_2, d_1, d_2$ in Eq. (\ref{vf3}) are uniquely
determined from six boundary conditions in Eq. (\ref{bc3}).
Electron and hole coefficients $u_i (k_j)$ and  $v_i (k_j)$
in wave functions Eq. (\ref{vf3}) are also found from Eq. (\ref{bogeq2}).
The excitation spectrum of a superconducting pnictide, corresponding to a fixed value of $k_y=0$ at zero angle of misorientation is shown in Fig. 3 (A),(C). The corresponding Fermi surface is shown in Fig. \ref{Fig3a}.
There are four intersection points with $k_{y}=0$.
At these points, $E(k_x)$ has minima shown in Figs. 3 (A) and (C).
 The existence of the four quasiparticle states in a pnictide with certain
sign of the group velocity follows from Figs. 3 (A) and (C).

The expression for the probability flow with fixed wave vector $k_y$ in the direction parallel to the $x$ axis follows from the BdG equations
on sites of the crystal lattice of a pnictide Eq. (\ref{bogeq2}) and has the following form:
\begin{gather}
J_p=\frac{2}{\hbar}((t_1+2t_3\cos k_y) {\rm Im}\{(\Psi^{\alpha}_{n+1})^*\Psi^{\alpha}_{n}-(\bar{\Psi}^{\alpha}_{n+1})^*\bar{\Psi}^{\alpha}_{n}\}\notag\\
+(t_2+2t_3\cos k_y) {\rm Im}\{(\Psi^{\beta}_{n+1})^*\Psi^{\beta}_{n}-(\bar{\Psi}^{\beta}_{n+1})^*\bar{\Psi}^{\beta}_{n}\}\notag\\
+4t_4\sin k_y {\rm Re}\{(\Psi^{\alpha}_{n+1})^*\Psi^{\beta}_{n}+(\Psi^{\beta}_{n+1})^*\Psi^{\alpha}_{n}\notag\\-(\bar{\Psi}^{\alpha}_{n+1})^*\bar{\Psi}^{\beta}_{n}-(\bar{\Psi}^{\beta}_{n+1})^*\bar{\Psi}^{\alpha}_{n}\}\notag\\
+2\Delta_0\cos k_y {\rm Im}\{(\Psi^{\alpha}_{n+1})^*\bar{\Psi}^{\alpha}_n\notag\\+(\bar{\Psi}^{\alpha}_{n+1})^*\Psi^{\alpha}_n+
(\Psi^{\beta}_{n+1})^*\bar{\Psi}^{\beta}_n+(\bar{\Psi}^{\beta}_{n+1})^*\Psi^{\beta}_n\}).\label{f4}
\end{gather}

\noindent One can show that the boundary conditions Eq. (\ref{bc3}) provide the conservation of the probability flow $J=J_p$ across the interface between a normal metal and a superconducting pnictide for each value of $k_y$. As well as in the case of previously considered 1D model (Eq. (\ref{f3})), the condition of flow conservation at the NS boundary, having the form of discrete sums (differences) on sites of the crystal lattice (Eqs. (\ref{f1}),(\ref{f4})), in the case of zero misorientation angle between crystallographic axes of a pnictide and the interface can be written in a quadratic form of the amplitudes of the probability to be in states with quasi-momenta $q_1, q_2, k_i, ~ i = 1 .. $4 multiplied by group velocities in these states.

The current through the two-dimensional microconstriction between a normal metal and a superconducting pnictide is determined by integration of Eq. (\ref{current}) over the transverse quasimomentum $k_y$: $I_p(V)=\eta_2\int dk_y I(V,k_y)$. In this case the probabilities $C, D$ of the quasiparticle propagation into a superconductor, given by Eqs. (\ref{current}) and (\ref{fout}), are determined by the sum of the scattering probabilities into individual bands: $C = C_1 + C_2, D = D_1 + D_2$. 
The coefficients  $A, B, C_1, C_2, D_1$, and $D_2$ in Eq. (\ref{fout}) are calculated from the boundary conditions Eq. (\ref{bc3}) and the expressions of
the probability flow in Eqs.  (\ref{f1}) and (\ref{f4}).
In the actual calculations, one should take into account that the original quasiparticle states 
should be normalized so that the probability flow in these states, described by the  Eqs. (\ref{f1}), (\ref{f4}), is equal to unity.

It is possible to demonstrate, that taking into account
oblique hopping between the boundary  $\gamma'_{1}, \gamma'_{2},\gamma''_{1},
\gamma''_{2}$ (see Fig. 2) allows to obtain  in quasiclassical limit  Ara\'{u}jo and Sacramento
boundary conditions \cite{sacr} only for the special case, when the following relations between hopping amplitudes are fulfilled simultaneously:

\begin{equation}
\left\{
\begin{aligned}
&\gamma_1=t_1, \\
&\gamma_2=t_2, \\
&\gamma_1'=\gamma_2'=(t_3-t_4), \\
&\gamma_1''=\gamma_2''=(t_3+t_4). \\
\end{aligned}
\right.\label{cond3}
\end{equation}

 \begin{figure}[h]
\centerline{\includegraphics[width=8cm]{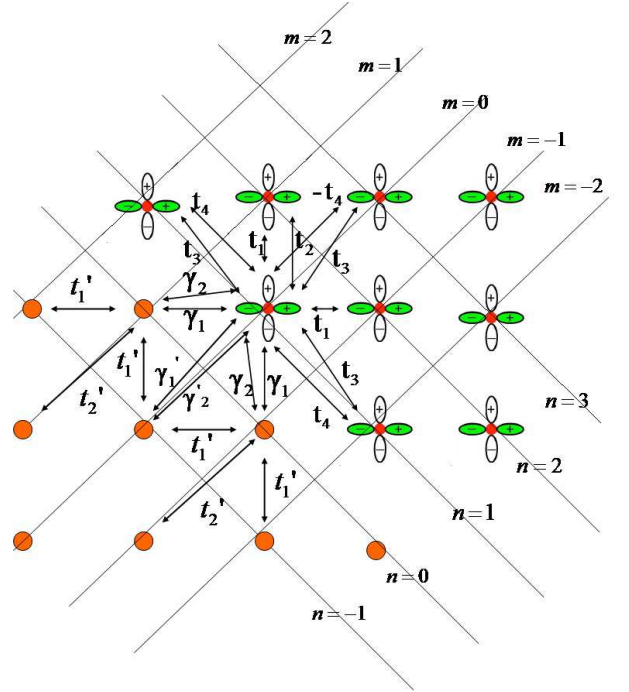}}
\caption{2D NS boundary. Angle between crystallographic axes of a pnictide and a normal metal is equal to $\pi/4$. The lower left region (orange circles) corresponds to the region of normal metal with hopping parameters $t_1',t_2'$, right region (sites with two $d$-orbitals) corresponds to the region of a superconducting pnictide with hopping parameters $t_1,t_2,t_3,t_4$. Boundary is described by hopping parameters $\gamma_1,\gamma_2,\gamma'_{1},\gamma'_{2}.$}
\label{Fig4}
\end{figure}

\subsection{\label{subsec2}Non-zero misorientation angle}

The proposed method allows one to consider the coherent electron transport in NS structures with non-zero misorientation angle as well.
It is necessary to note that the microscopic calculation of the conductance for a non-zero misorientation angle of a pnictide crystal with respect to the
boundary is presented here for the first time. Previous
phenomenological approaches \cite{sacr, lind} don't allow one to carry out such calculations.
In considering the electron transport across NS contact with a nonzero misorientation angle, it is necessary to take into account hopping at the
two adjacent atomic layers of a pnictide (Fig.\ref{Fig4}).
BdG equations in the considered case  corresponding to $s_{\pm}$ symmetry of the pair potential in pnictides are given in the Appendix, see Eq. (\ref{bogeq3}).
Hopping across the NS boundary for non-zero misorientation angle is described by a larger number of parameters, rather than at zero misorientation angle between crystallographic axes of a pnictide and the interface (see Fig.\ref{Fig4}). In addition to hopping parameters $\gamma_1$ and
$\gamma_2$,  we should use additional parameters of hopping across the boundary $\gamma'_{1}$ and
$\gamma'_{2}$. These parameters of hopping across the boundary take into account connection of orbitals from the last atom layer of a pnictide with the penultimate from the boundary atom layer of the normal metal.  Taking into account these processes is necessary due to the breaking at the boundary of the diagonal bonds in the crystal lattice of a pnictide for non-zero angle of misorientation (see Fig.\ref{Fig4}). Also in the normal metal together with the nearest neighbor hopping $t_1 '$ we need to consider the diagonal hopping $t_2' $  in square lattice.

 The wave functions Eq. (\ref{vf4}) and the relation for probability flow (\ref{fl2}) take into account
not only the electron transport in two energy bands, but also in two valleys in these bands (see Fig. \ref{Fig3}B,D). It is known from the physics of semiconductors, that interference of states in the valleys is possible.\cite{ando}
This interference leads to the fact that the condition of flow conservation at the boundary of the NS contact, having the form of discrete sums (differences) on sites of the crystal lattice, in the case of nonzero misorientation angle between crystallographic axes of a pnictide and the interface can not be written in a quadratic form of the amplitudes of the probability to be in states with quasimomentum $q_1, q_2, k_i, ~ i = 1 .. $8, multiplied by group velocities in these states.

\subsection{\label{subsec3}Numerical results}

 \begin{figure}[b]
\centerline{\includegraphics[width=9cm]{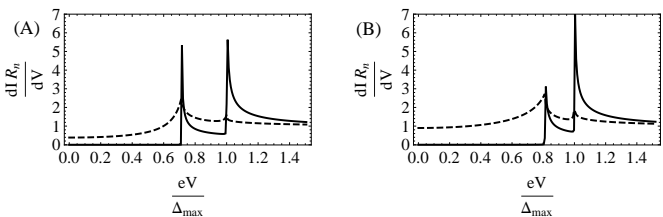}}
\caption{Angle resolved conductance without misorientation for (A) $s_{\pm}$-model and (B) $s_{++}$-model.
Value of the quasimomentum, parallel to the interface, $k_y=0.01$.
The values of hopping parameters at the interface are chosen as $\gamma_1 = 0.1, \gamma_2 = 0.14$ (eV) (dashed line),
and $\gamma_1 = 0.009, \gamma_2 = 0.005$ (eV) (solid lines).
}
\label{Fig5}
\end{figure}

 \begin{figure}[t]
\centerline{\includegraphics[width=9cm]{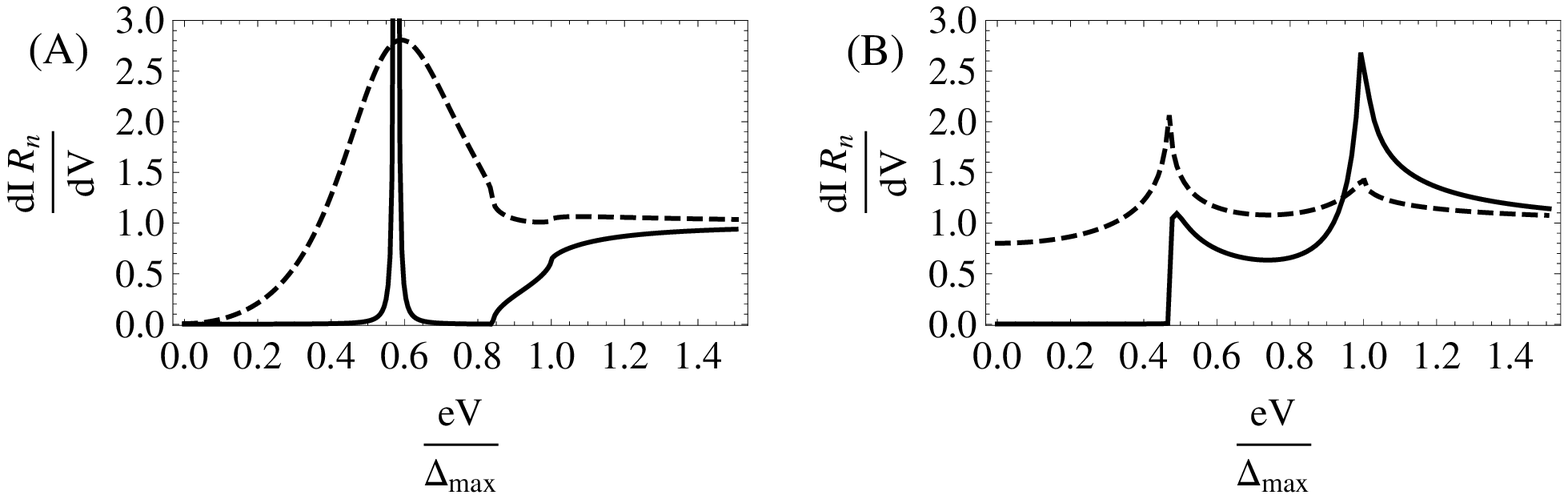}}
\caption{Same as Fig. \ref{Fig5} but with $k_y=3\pi/4$.}
\label{Fig6}
\end{figure}

 \begin{figure}[t]
\centerline{\includegraphics[width=9cm]{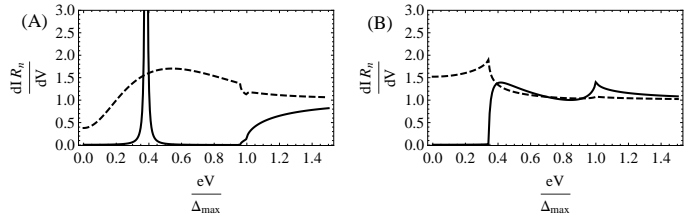}}
\caption{Same as Fig. \ref{Fig5} but with $k_y=5\pi/8$}
\label{Fig7}
\end{figure}

\begin{figure}[t]
\centerline{\includegraphics[width=9cm]{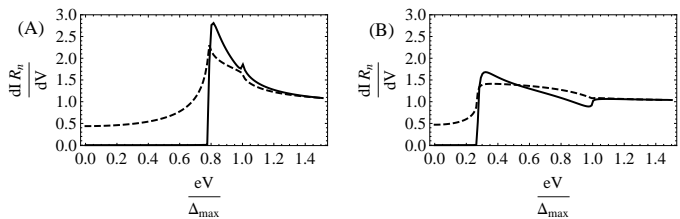}}
\caption{
Angle resolved conductance with misorientation angle $\pi/4$ for (A) $s_{\pm}$-model and (B) $s_{++}$-model.
Value of the quasimomentum, parallel to the interface, $k_y=0$.
The values of hopping parameters at the interface are chosen as
$\gamma_1 = 0.1, \gamma_2 = 0.14, \gamma'_1 = 0.2, \gamma'_2 = 0.06$ (eV) (dashed line),
and $\gamma_1 = 0.009, \gamma_2 = 0.005, \gamma'_1 = 0.02, \gamma'_2 = 0.01$ (eV) (solid lines).
}
\label{Fig8}
\end{figure}

 \begin{figure}[t]
\centerline{\includegraphics[width=9cm]{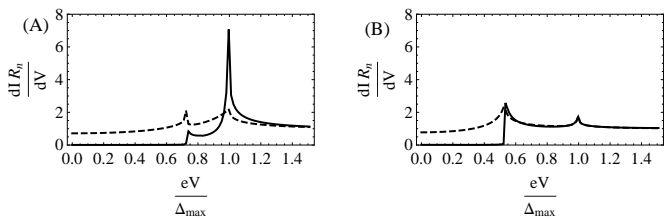}}
\caption{Same as Fig. \ref{Fig8} but with $k_y=\pi/3$.}
\label{Fig9}
\end{figure}

Here we will show the results of numerical calculations of angle-resolved
conductance(dI/dV) as a function of bias voltage $V$
in normal metal / superconducting pnictide junctions.
We use the following values of hopping parameters and chemical potential in a pnictide:
$t_1 = -0.1051$, $t_2 = 0.1472$, $t_3 = -0.1909$, $t_4 = -0.0874$ and $\mu_{S}=-0.081$ (eV), according to Ref. \onlinecite{mor}.
We assume the momentum dependence of the pair potential
in the $s_\pm$ model as $\Delta_{\pm}(k)=4\Delta_0\cos{k_x}\cos{k_y}$ with $\Delta_0=0.008$ (eV),
and $s_{++}$ model in the form $\Delta=2\Delta_0(\cos k_x+\cos k_y)+\Delta_1$ with $\Delta_0=0.002, \Delta_1=0.0042$ (eV).
In Figs. \ref{Fig5} to \ref{Fig7},
the magnitudes of
tunneling conductance normalized by their values in the normal state are shown
for $s_{\pm}$ and $s_{++}$ models with zero misorientation angle.
The hopping parameters and chemical potential in normal metal are $t_1' = 0.3$, $t_2'=0$, $\mu_{N}=0.2$.
For the hopping parameters at the interface,
we choose two cases with
$\gamma_1 = 0.009,
\gamma_2 = 0.005$ (eV) (low transmissivity) and $\gamma_1 = 0.1,
\gamma_2 = 0.14$ (eV) (high transmissivity).
Calculated charge conductance $dI/dV$ for low and high transparent junctions
with $k_y=0.01$ is shown in Fig. \ref{Fig5}.
The horizontal axis represents $eV$ normalized by $\Delta_{max}$,
where $\Delta_{max}$ is the maximum of two gaps for fixed $k_{y}$.
One can clearly see two gap features
reflecting the presence of two kinds of Fermi surfaces (see Fig. \ref{Fig3})
in both $s_{\pm}$- Fig.\ref{Fig5}(A)
and $s_{++}$- cases Fig.\ref{Fig5}(B).
In the case of $k_y\sim 0$, the interorbital hopping $t_4$ is absent.
Therefore, the obtained conductance can be represented by
a simple summation of the individual orbital's contributions.
On the other hand, in the case of $s_{\pm}$-wave with low transmissive interface, a sharp subgap peak appears in the energy gap as shown in Fig.\ref{Fig6}(A) and Fig.\ref{Fig7}(A), respectively.
These energy structures do not
correspond to the density of states in the bulk.
Since $dI/dV$ corresponds to the energy spectrum of local density of states in
the low transmissivity, we can conclude
that these subgap structures originate from the surface Andreev bound states at finite energies.
The bound states disappear in the case of high transparency of the interface.
On the other hand, as is seen from Fig.\ref{Fig6}(B) and Fig.\ref{Fig7}(B), these features are not present in the case of $s_{++}$-wave when the signs of pair potentials in different bands are the same.
From these results, we can conclude that the surface Andreev bound states are formed in the $s_{\pm}$ case due to the sign change of pair potential and the interorbital hopping $t_4$.
Note that a sharp subgap peak in the angle-resolved conductance discussed here should be broadened after summation over $k_y$ is made, as calculated by Onari {\it et al} within different model.\cite{Onari2009}

Next, we calculate the case with finite misorientation angle $\pi/4$.
In this case, we shall introduce additional hopping coefficients in a normal metal ($t_2'$) and at the interface ($\gamma'_{1}$, $\gamma'_{2}$) corresponding to the direction perpendicular to the interface.
We choose $t_2'=0.01$ (eV), $\gamma_{1}' = 0.02, \gamma'_{2} = 0.01$ (eV) (low transmissivity case) and $\gamma'_{1} = 0.2,
\gamma'_{2} = 0.06$ (eV) (high transmissivity case).
Other hopping parameters are the same as in the case with zero misorientation.
In Figs. \ref{Fig8} and \ref{Fig9},
we show the angle resolved conductances for $k_y=0$ and $\pi/3$, respectively.
In the case of $s_{\pm}$ model, one can see the two gap features for both $k_y=0$ and $\pi/3$ as shown in Figs. \ref{Fig8}(A) and \ref{Fig9}(A).
Subgap peaks are absent even in low transparent junctions, in contrast to the case with zero misorientation angle.
This is because no sign change at fixed $k_y$ values occurs
in the case when misorientation angle equals $\pi/4$ (Fig. \ref{Fig3a}(B)).
For the same reason, gap structure without subgap peaks appears also in the case of $s_{++}$ model.

Let us summarize the results of the conductance of normal metal / superconducting pnictide junctions.
In the case of $s_{++}$ model, only the two-gap structure without subgap peaks
appears for any misorientation angle and any value of $k_y$.
On the other hand, in the case of $s_{\pm}$ model with low transparent interface,
subgap peaks appear for zero misorientation angle and finite $k_y$.
These subgap peaks originate from sign change of the pair potential
at fixed $k_y$ values in the presence of the interorbital hopping.


\section{Conclusion}

In this paper, we have presented consistent tight-binding model for the coherent electronic transport in the contact between a normal metal and a superconductor.
Based on a tight-binding model beyond effective mass
approximation, we have derived boundary conditions on a wave function at a contact between
a normal metal and a superconductor with unconventional pairing symmetry.
We have extended the previous tight-binding approach used in semiconducting heterostructures \cite{krom} to the case of superconducting junctions.
The obtained boundary conditions contain real space information only without any momentum derivatives, and they have clear physical meaning.
These conditions provide current conservation and enable one
to formulate consistent approach for tunneling spectroscopy of
superconductors with  complex nonparabolic energy spectrum,
including  multiband electronic structure and unconventional symmetry of superconducting pairing.
We have shown that application of this theory to single-band superconductor junctions allows one to
reproduce the preexisting conductance formula \onlinecite{tanaka1}.
Based on the derived boundary conditions, we have calculated conductance in normal metal / superconducting pnictide junctions
for different misorientation angles between the interface and the crystallographic axes of a pnictide.
The present approach provides the basis for tunneling spectroscopy of multi-orbital superconductors.
Moreover, this approach is suitable for a consistent description of electronic transport in structures with surface states
described by Majorana fermions in topological superconductors,
\cite{tanaka12,fu08,fu09,akhmerov09,law09,
linder10,tanaka09,sato09,yokoyama12,Alicea1,oreg10,
lutchyn10,ioselevich11,Yamakage2,TanakaRapid,Yada1,Yada2}
which would be the subject of our future study.

\begin{acknowledgments}
We gratefully acknowledge M.Yu. Kupriyanov, I.I. Mazin, A. S. Melnikov and S. Onari for valuable discussions.
This work was supported in part by a Grant-in Aid for Scientific Research from MEXT of Japan, "Topological
Quantum Phenomena" Grants No. 22103005 and No. 20654030 (Y.T.), RFBR Grant ¹ 11-02-12084-ofi-m-2011,
Dutch Foundation for Fundamental Research on Matter (FOM) and by EU-Japan program "IRON SEA".
\end{acknowledgments}


\appendix

\section{Derivation of equations of two-dimensional model  \label{A}}

Bogoliubov-de Gennes equations on sites of the Fe crystal lattice in the $x-y$ plane of a pnictide for the case of zero misorientation angle of the crystallographic axes of pnictides with respect to the interface have the following form:
\begin{widetext}
\begin{equation}
\left\{
\begin{aligned}
&t_1( \Psi^{\alpha}_{n+1,m}+ \Psi^{\alpha}_{n-1,m})+t_2( \Psi^{\alpha}_{n,m+1}+ \Psi^{\alpha}_{n,m-1})\\&+t_3( \Psi^{\alpha}_{n+1,m+1}+ \Psi^{\alpha}_{n-1,m-1}+ \Psi^{\alpha}_{n+1,m-1}\\&+\Psi^{\alpha}_{n-1,m+1})+t_4( -\Psi^{\beta}_{n+1,m+1}- \Psi^{\beta}_{n-1,m-1}\\&+ \Psi^{\beta}_{n+1,m-1}+ \Psi^{\beta}_{n-1,m+1})-\mu_{S}\Psi^{\alpha}_{n,m} +\Delta_0( \bar{\Psi}^{\alpha}_{n+1,m+1}\\&+\bar{\Psi}^{\alpha}_{n-1,m-1}+\bar{\Psi}^{\alpha}_{n+1,m-1}+ \bar{\Psi}^{\alpha}_{n-1,m+1})=\varepsilon\Psi^{\alpha}_{n,m}, \\
&t_2( \Psi^{\beta}_{n+1,m}+ \Psi^{\beta}_{n-1,m})+t_1( \Psi^{\beta}_{n,m+1}+ \Psi^{\beta}_{n,m-1})\\&+t_3( \Psi^{\beta}_{n+1,m+1}+ \Psi^{\beta}_{n-1,m-1}+ \Psi^{\beta}_{n+1,m-1}\\&+ \Psi^{\beta}_{n-1,m+1})+t_4( -\Psi^{\alpha}_{n+1,m+1}- \Psi^{\alpha}_{n-1,m-1}\\&+\Psi^{\alpha}_{n+1,m-1}+ \Psi^{\alpha}_{n-1,m+1})-\mu_{S}\Psi^{\beta}_{n,m} +\Delta_0( \bar{\Psi}^{\beta}_{n+1,m+1}\\&+\bar{\Psi}^{\beta}_{n-1,m-1}+\bar{\Psi}^{\beta}_{n+1,m-1}+ \bar{\Psi}^{\beta}_{n-1,m+1})=\varepsilon\Psi^{\beta}_{n,m}, \\
&t_1( \bar{\Psi}^{\alpha}_{n+1,m}+ \bar{\Psi}^{\alpha}_{n-1,m})+t_2(\bar{\Psi}^{\alpha}_{n,m+1}+ \bar{\Psi}^{\alpha}_{n,m-1})\\&+t_3( \bar{\Psi}^{\alpha}_{n+1,m+1}+ \bar{\Psi}^{\alpha}_{n-1,m-1}+ \bar{\Psi}^{\alpha}_{n+1,m-1}\\&+\bar{\Psi}^{\alpha}_{n-1,m+1})+t_4( -\bar{\Psi}^{\beta}_{n+1,m+1}- \bar{\Psi}^{\beta}_{n-1,m-1}\\&+\bar{\Psi}^{\beta}_{n+1,m-1}+\bar{\Psi}^{\beta}_{n-1,m+1}) -\mu_{S}\bar{\Psi}^{\alpha}_{n,m} -\Delta_0( \Psi^{\alpha}_{n+1,m+1}\\&+\Psi^{\alpha}_{n-1,m-1}+\Psi^{\alpha}_{n+1,m-1}+ \Psi^{\alpha}_{n-1,m+1})=-\varepsilon\bar{\Psi}^{\alpha}_{n,m},\\
&t_2( \bar{\Psi}^{\beta}_{n+1,m}+ \bar{\Psi}^{\beta}_{n-1,m})+t_1( \bar{\Psi}^{\beta}_{n,m+1}+ \bar{\Psi}^{\beta}_{n,m-1})\\&+t_3( \bar{\Psi}^{\beta}_{n+1,m+1}+ \bar{\Psi}^{\beta}_{n-1,m-1}+ \bar{\Psi}^{\beta}_{n+1,m-1}\\&+\bar{\Psi}^{\beta}_{n-1,m+1})+t_4( -\bar{\Psi}^{\alpha}_{n+1,m+1}- \bar{\Psi}^{\alpha}_{n-1,m-1}\\&+\bar{\Psi}^{\alpha}_{n+1,m-1}+ \bar{\Psi}^{\alpha}_{n-1,m+1}) -\mu_{S}\bar{\Psi}^{\beta}_{n,m} -\Delta_0( \Psi^{\beta}_{n+1,m+1}\\& +\Psi^{\beta}_{n-1,m-1}+\Psi^{\beta}_{n+1,m-1}+ \Psi^{\beta}_{n-1,m+1})=-\varepsilon\bar{\Psi}^{\beta}_{n,m},
\end{aligned}
\right.\label{bogeq2}
\end{equation}
\end{widetext}
\noindent where $t_i,~i=1..4$, are hopping amplitudes between orbitals on sites in a pnictide in the two-orbital model.\cite{rag} The value of $\Delta_0 $ is the amplitude of the anisotropic pair potential corresponding to the considered $s_{\pm}$ superconducting pairing model: $\Delta_{\pm} (k) = 4\Delta_0 \cos {k_x} \cos {k_y}$,\cite{kor} $k_y,k_x$ are parallel and perpendicular to the interface components of quasimomentum respectively.  The wave functions of a superconducting pnictide have the upper orbital index $\alpha (\beta)$: $\Psi^{\alpha (\beta)}_i$, corresponding to $d_{xz} (d_{yz})$ orbital respectively. The subscripts $n,m$ of the wave function $\Psi^{\alpha (\beta)}_ {n,m}$ of a pnictide describe the coordinates of sites of the crystal lattice (Fig. 2). As well as in the considered above 1D-model $\Psi^{\alpha (\beta)} _ {n,m}$ in Eq. (\ref{bogeq2}) describe the electron states, and $\bar{\Psi}^{\alpha (\beta)}_{n,m}$ - hole states.

For misorientation angle $\pi/4$ between crystallographic axes of a pnictide and the interface (Fig.\ref{Fig4}), Bogoliubov-de Gennes equations  on sites of the Fe crystal lattice in the $x-y$ plane of a pnictide differ from Eq. (\ref{bogeq2}) and have the following form:

\clearpage

\begin{equation}
\left\{
\begin{aligned}
&-\mu_{S}\Psi^{\alpha}_{n,m}+t_1( \Psi^{\alpha}_{n+1,m-1}+ \Psi^{\alpha}_{n-1,m+1})\\&+t_2( \Psi^{\alpha}_{n+1,m+1}+ \Psi^{\alpha}_{n-1,m-1})\\&+t_3( \Psi^{\alpha}_{n+2,m}+ \Psi^{\alpha}_{n-2,m}+ \Psi^{\alpha}_{n,m-2}+ \Psi^{\alpha}_{n,m+2})\\&+t_4( -\Psi^{\beta}_{n+2,m}- \Psi^{\beta}_{n-2,m}+ \Psi^{\beta}_{n,m-2}+ \Psi^{\beta}_{n,m+2})\\&+\Delta_0( \bar{\Psi}^{\alpha}_{n+2,m}+ \bar{\Psi}^{\alpha}_{n-2,m}+\bar{\Psi}^{\alpha}_{n,m-2}+ \bar{\Psi}^{\alpha}_{n,m+2})\\& =\varepsilon\Psi^{\alpha}_{n,m}, \\
&-\mu_{S}\Psi^{\beta}_{n,m}+t_2( \Psi^{\beta}_{n+1,m-1}+ \Psi^{\beta}_{n-1,m+1})\\&+t_1( \Psi^{\beta}_{n+1,m+1}+ \Psi^{\beta}_{n-1,m-1})\\&+t_3( \Psi^{\beta}_{n+2,m}+ \Psi^{\beta}_{n-2,m}+ \Psi^{\beta}_{n,m-2}+ \Psi^{\beta}_{n,m+2})\\&+t_4( -\Psi^{\alpha}_{n+2,m}- \Psi^{\alpha}_{n-2,m}+ \Psi^{\alpha}_{n,m-2}+ \Psi^{\alpha}_{n,m+2})\\&+\Delta_0( \bar{\Psi}^{\beta}_{n+2,m}+ \bar{\Psi}^{\beta}_{n-2,m}+\bar{\Psi}^{\beta}_{n,m-2}+ \bar{\Psi}^{\beta}_{n,m+2})\\& =\varepsilon\Psi^{\beta}_{n,m}, \\
&-\mu_{S}\bar{\Psi}^{\alpha}_{n,m}+t_1( \bar{\Psi}^{\alpha}_{n+1,m-1}+ \bar{\Psi}^{\alpha}_{n-1,m+1})\\&+t_2(\bar{\Psi}^{\alpha}_{n+1,m+1}+ \bar{\Psi}^{\alpha}_{n-1,m-1})\\&+t_3( \bar{\Psi}^{\alpha}_{n+2,m}+ \bar{\Psi}^{\alpha}_{n-2,m}+ \bar{\Psi}^{\alpha}_{n,m-2}+ \bar{\Psi}^{\alpha}_{n,m+2})\\&+t_4( -\bar{\Psi}^{\beta}_{n+2,m}- \bar{\Psi}^{\beta}_{n-2,m}+ \bar{\Psi}^{\beta}_{n,m-2}+\bar{\Psi}^{\beta}_{n,m+2}) \\&-\Delta_0( \Psi^{\alpha}_{n+2,m}+ \Psi^{\alpha}_{n-2,m}+\Psi^{\alpha}_{n,m-2}+ \Psi^{\alpha}_{n,m+2})\\& =-\varepsilon\bar{\Psi}^{\alpha}_{n,m},\\
&-\mu_{S}\bar{\Psi}^{\beta}_{n,m}+t_2( \bar{\Psi}^{\beta}_{n+1,m-1}+ \bar{\Psi}^{\beta}_{n-1,m+1})\\&+t_2( \bar{\Psi}^{\beta}_{n+1,m+1}+ \bar{\Psi}^{\beta}_{n-1,m-1})\\&+t_3( \bar{\Psi}^{\beta}_{n+2,m}+ \bar{\Psi}^{\beta}_{n-2,m}+ \bar{\Psi}^{\beta}_{n,m-2}+ \bar{\Psi}^{\beta}_{n,m+2})\\&+t_4( -\bar{\Psi}^{\alpha}_{n+2,m}- \bar{\Psi}^{\alpha}_{n-2,m}+ \bar{\Psi}^{\alpha}_{n,m-2}+ \bar{\Psi}^{\alpha}_{n,m+2})\\&-\Delta_0( \Psi^{\beta}_{n+2,m}+ \Psi^{\beta}_{n-2,m}+\Psi^{\beta}_{n,m-2}+ \Psi^{\beta}_{n,m+2}) \\&=-\varepsilon\bar{\Psi}^{\beta}_{n,m}.
\end{aligned}
\right.\label{bogeq3}
\end{equation}

The boundary conditions for the contact between a normal metal and a pnictide, considered in the framework of the two-orbital model, for misorientation angle $\pi/4$ between crystallographic axes of a pnictide and the interface have the following form:

\begin{gather}
\left\{
\begin{aligned}
&t'_1\Phi_1(e^{ik_yl}+e^{-ik_yl})+t'_2\Phi_2
=\Psi^{\alpha}_1(\gamma_1e^{ik_yl}
+\gamma_2e^{-ik_yl})\\&
+\Psi^{\beta}_1(\gamma_1e^{-ik_yl}+\gamma_2e^{ik_yl})
+\gamma'_1\Psi^{\alpha}_2+\gamma'_2\Psi^{\beta}_2,\\
&t'_1\bar{\Phi}_1(e^{ik_yl}+e^{-ik_yl})+t'_2\bar{\Phi}_2=
\bar{\Psi}^{\alpha}_1(\gamma_1e^{ik_yl}+\gamma_2e^{-ik_yl})\\&
+\bar{\Psi}^{\beta}_1(\gamma_1e^{-ik_yl}+\gamma_2e^{-ik_yl})
+\gamma'_1\bar{\Psi}^{\alpha}_2+\gamma'_2\bar{\Psi}^{\beta}_2,\\
&\Phi_0(\gamma_1
e^{ik_yl}+\gamma_2
e^{-ik_yl})+\gamma'_1\Phi_{-1}=t_1\Psi^{\alpha}_{0}e^{ik_yl}\\&+t_2\Psi^{\alpha}_{0}e^{-ik_yl}+t_3\Psi^{\alpha}_{-1}-t_4\Psi^{\beta}_{-1}+\Delta_0
 \bar{\Psi}^{\alpha}_{-1},\\
&\bar{\Phi}_0(\gamma_1
e^{ik_yl}+\gamma_2
e^{-ik_yl})+\gamma'_1\bar{\Phi}_{-1}=t_1\bar{\Psi}^{\alpha}_{0}e^{ik_yl}\\&+t_2\bar{\Psi}^{\alpha}_{0}e^{-ik_yl}+t_3\bar{\Psi}^{\alpha}_{-1}-t_4\bar{\Psi}^{\beta}_{-1}-\Delta_0
\Psi^{\alpha}_{-1},\\
&\Phi_0(\gamma_1
e^{-ik_yl}+\gamma_2
e^{ik_yl})+\gamma'_2\Phi_{-1}=t_1\Psi^{\beta}_{0}e^{ik_yl}\\&+t_2\Psi^{\beta}_{0}e^{-ik_yl}+t_3\Psi^{\beta}_{-1}-t_4\Psi^{\alpha}_{-1}+\Delta_0
 \bar{\Psi}^{\beta}_{-1},\\
 &\bar{\Phi}_0(\gamma_1
e^{-ik_yl}+\gamma_2
e^{ik_yl})+\gamma'_2\bar{\Phi}_{-1}=t_1\bar{\Psi}^{\beta}_{0}e^{ik_yl}\\&+t_2\bar{\Psi}^{\beta}_{0}e^{-ik_yl}+t_3\bar{\Psi}^{\beta}_{-1}-t_4\bar{\Psi}^{\alpha}_{-1}-\Delta_0
\Psi^{\beta}_{-1},\\
&\gamma'_1\Phi_0=t_3\Psi^{\alpha}_0-t_4\Psi^{\beta}_0+\Delta_0\bar{\Psi}^{\alpha}_0,\\
&\gamma'_1\bar{\Phi}_0=t_3\bar{\Psi}^{\alpha}_0-t_4\bar{\Psi}^{\beta}_0-\Delta_0\Psi^{\alpha}_0,\\
&\gamma'_2\Phi_0=t_3\Psi^{\beta}_0-t_4\Psi^{\alpha}_0+\Delta_0\bar{\Psi}^{\beta}_0,\\
&\gamma'_2\bar{\Phi}_0=t_3\bar{\Psi}^{\beta}_0-t_4\bar{\Psi}^{\alpha}_0-\Delta_0\Psi^{\beta}_0,\\
&t'_2\Phi_1=\gamma'_1\Psi^{\alpha}_1+\gamma'_2\Psi^{\beta}_1,\\
&t'_2\bar{\Phi}_1=\gamma'_1\bar{\Psi}^{\alpha}_1+\gamma'_2\bar{\Psi}^{\beta}_1.
\end{aligned}
\right.\label{bc4}
\end{gather}

\noindent As in the previously considered case of boundary conditions for zero misorientation angle Eq. (\ref{bc3}), due to translational symmetry in the direction parallel to the boundary, in electron (hole) wave functions $\Psi^{\alpha (\beta)}_{n,m}$ ($\bar{\Psi}^{\alpha (\beta)}_{n,m}$) second subscript $(m)$ corresponding to the coordinate of an atom in a direction parallel to the boundary is omitted.

The wave functions in a normal metal / superconducting pnictide contact in the case of misorientation angle $\pi/4$ between crystallographic axes of a pnictidea and the interface are defined by eight plane waves with amplitudes $a_1,b_1,a_2,b_2,c_1, c_2, d_1, d_2, f_1, f_2, g_1, g_2$. Here the coefficients $a_1, b_1,a_2,b_2$ describe Andreev and normal reflected waves, while $c_1, c_2, d_1, d_2, f_1, f_2, g_1, g_2$ describe eight waves transmitted into a two-band superconducting pnictide:

\begin{equation}
\left\{
\begin{aligned}
&\Phi_n=\exp(iq_1nl)+b_1~\exp(-iq_1nl)+b_2~\exp(-iq_2nl),\\
&\bar{\Phi}_n=a_1~\exp(iq_3nl)+a_2~\exp(iq_4nl),\\
&\Psi^{\alpha}_n=c_1u_1(k_1)\exp(ik_1nl)+c_2u_1(k_2)\exp(ik_2nl)\\&+d_1u_1(k_3)\exp(ik_3nl)+d_2u_1(k_4)\exp(ik_4nl)
\\&+f_1u_1(k_5)\exp(ik_5nl)+f_2u_1(k_6)\exp(ik_6nl)\\&+g_1u_1(k_7)\exp(ik_7nl)+g_2u_1(k_8)\exp(ik_8nl),\\
&\Psi^{\beta}_n=c_1u_2(k_1)\exp(ik_1nl)+c_2u_2(k_2)\exp(ik_2nl)\\&+d_1u_2(k_3)\exp(ik_3nl)+d_2u_2(k_4)\exp(ik_4nl)
\\&+f_1u_2(k_5)\exp(ik_5nl)+f_2u_2(k_6)\exp(ik_6nl)\\&+g_1u_2(k_7)\exp(ik_7nl)+g_2u_2(k_8)\exp(ik_8nl),\\
&\bar{\Psi}^{\alpha}_n=c_1v_1(k_1)\exp(ik_1nl)+c_2v_1(k_2)\exp(ik_2nl)\\&+d_1v_1(k_3)\exp(ik_3nl)+d_2v_1(k_4)\exp(ik_4nl)
\\&+f_1v_1(k_5)\exp(ik_5nl)+f_2v_1(k_6)\exp(ik_6nl)\\&+g_1v_1(k_7)\exp(ik_7nl)+g_2v_1(k_8)\exp(ik_8nl),\\
&\bar{\Psi}^{\beta}_n=c_1v_2(k_1)\exp(ik_1nl)+c_2v_2(k_2)\exp(ik_2nl)\\&+d_1v_2(k_3)\exp(ik_3nl)+d_2v_2(k_4)\exp(ik_4nl)
\\&+f_1v_2(k_5)\exp(ik_5nl)+f_2v_2(k_6)\exp(ik_6nl)\\&+g_1v_2(k_7)\exp(ik_7nl)+g_2v_2(k_8)\exp(ik_8nl).
\end{aligned}
\right.\label{vf4}
\end{equation}

\noindent Four transmitted waves with amplitudes $c_1, c_2, d_1, d_2$ correspond to the lower band, depicted by black solid line on Fig. \ref{Fig3}(B),(D). These four waves are propagating waves except the energy range lower than the superconducting gap $\Delta_0$. Four plane waves with amplitudes $f_1, f_2, g_1, g_2$ correspond to the upper band, depicted by red dashed line on Fig. \ref{Fig3}(B),(D). These four waves are evanescent waves on the scale of pair potential  $\Delta_0$.

Expression for the probability flow in the case of misorientation angle between crystallographic axes of a pnictide and the interface equal to $\pi/4$
differs from the corresponding relation for the case of zero misorientation angle Eq. (\ref{f4}) and
has the following form:

\begin{gather}
J=\frac{2}{\hbar}(t_1{\rm Im}\{(\Psi^{\alpha}_{n+1})^*\Psi^{\alpha}_{n}e^{ik_yl}\}\notag\\+t_2{\rm Im}\{(\Psi^{\alpha}_{n+1})^*\Psi^{\alpha}_{n}e^{-ik_yl}\}\notag\\+t_3{\rm Im}\{(\Psi^{\alpha}_{n+1})^*\Psi^{\alpha}_{n-1}+(\Psi^{\alpha}_{n+2})^*\Psi^{\alpha}_{n}\}\notag\\+
t_1{\rm Im}\{(\Psi^{\beta}_{n+1})^*\Psi^{\beta}_{n}e^{-ik_yl}\}+t_2{\rm Im}\{(\Psi^{\beta}_{n+1})^*\Psi^{\beta}_{n}e^{ik_yl}\}\notag\\+t_3{\rm Im}\{(\Psi^{\beta}_{n+1})^*\Psi^{\beta}_{n-1}+(\Psi^{\beta}_{n+2})^*\Psi^{\beta}_{n}\}\notag\\
-t4({\rm Im}\{(\Psi^{\alpha}_{n+1})^*\Psi^{\beta}_{n-1}\}+{\rm Im}\{(\Psi^{\beta}_{n+1})^*\Psi^{\alpha}_{n-1}\}
\notag\\+{\rm Im}\{(\Psi^{\alpha}_{n+2})^*\Psi^{\beta}_{n}\}+{\rm Im}\{(\Psi^{\beta}_{n+2})^*\Psi^{\alpha}_{n}\})\notag\\-
t_1{\rm Im}\{(\bar{\Psi}^{\alpha}_{n+1})^*\bar{\Psi}^{\alpha}_{n}e^{ik_yl}\}-t_2{\rm Im}\{(\bar{\Psi}^{\alpha}_{n+1})^*\bar{\Psi}^{\alpha}_{n}e^{-ik_yl}\}\notag\\-t_3{\rm Im}\{(\bar{\Psi}^{\alpha}_{n+1})^*\bar{\Psi}^{\alpha}_{n-1}+(\bar{\Psi}^{\alpha}_{n+2})^*\bar{\Psi}^{\alpha}_{n}\}\notag\\
-t_1{\rm Im}\{(\bar{\Psi}^{\beta}_{n+1})^*\bar{\Psi}^{\beta}_{n}e^{-ik_yl}\}-t_2{\rm Im}\{(\bar{\Psi}^{\beta}_{n+1})^*\bar{\Psi}^{\beta}_{n}e^{ik_yl}\notag\\-t_3{\rm Im}\{(\bar{\Psi}^{\beta}_{n+1})^*\bar{\Psi}^{\beta}_{n-1}+(\bar{\Psi}^{\beta}_{n+2})^*\bar{\Psi}^{\beta}_{n}\}\notag\\
+t4({\rm Im}\{(\bar{\Psi}^{\alpha}_{n+1})^*\bar{\Psi}^{\beta}_{n-1}\}+{\rm Im}\{(\bar{\Psi}^{\beta}_{n+1})^*\bar{\Psi}^{\alpha}_{n-1}\}
\notag\\+{\rm Im}\{(\bar{\Psi}^{\alpha}_{n+2})^*\bar{\Psi}^{\beta}_{n}\}+{\rm Im}\{(\bar{\Psi}^{\beta}_{n+2})^*\bar{\Psi}^{\alpha}_{n}\})\notag\\+
\Delta_0{\rm Im}\{(\Psi^{\alpha}_{n+1})^*\bar{\Psi}^{\alpha}_{n-1}+(\bar{\Psi}^{\alpha}_{n+1})^*\Psi^{\alpha}_{n-1}\notag\\+
(\Psi^{\beta}_{n+1})^*\bar{\Psi}^{\beta}_{n-1}+(\bar{\Psi}^{\beta}_{n+1})^*\Psi^{\beta}_{n-1}+
(\Psi^{\alpha}_{n+2})^*\bar{\Psi}^{\alpha}_{n}\notag\\+(\bar{\Psi}^{\alpha}_{n+2})^*\Psi^{\alpha}_{n}+
(\Psi^{\beta}_{n+2})^*\bar{\Psi}^{\beta}_{n}+(\bar{\Psi}^{\beta}_{n+2})^*\Psi^{\beta}_{n}\}).
\label{fl2}
\end{gather}

\bibliography{biblio}

\end{document}